\documentstyle[12pt]{article}

\begin{document}

\begin{titlepage}
\begin{flushright} {INRNE\\ Th-02/94} \end{flushright}

\vskip 3.6truecm

\begin{center}
{\large \bf
SOME FUNCTIONAL SOLUTIONS OF \\
THE YANG-BAXTER EQUATION }
\end{center}

\vskip 1.0cm

\begin{center}
D. Ts. Stoyanov$^{\star\ddag}$ \\
\vskip 0.2cm
{\it Institute for Nuclear Research and Nuclear Energy \\
Boul.~Tsarigradsko chaussee 72, Sofia 1784, Bulgaria}
\vskip 1cm
28th April 1994
\end{center}

\vskip 2.3cm

\rm
{\bf Abstract. }
A general functional definition of the infinite
dimensional quantum $R$-matrix satisfying the Yang-Baxter
equation is given.
A procedure for the extracting a finite dimensional
$R$-matrix from  the  general  definition  is  demonstrated  in  a
particular case when the group $SU(2)$ takes place.
\vfill

\begin{flushleft}
\rule{5.1 in}{.007 in}\\
$^{\star}$ {\small Supported in part by BNSF under contract Ph-20-91}\\
$^{\ddag}$ {\small Bitnet address: DSTOYAN@BGEARN \\ }
\end{flushleft}

\end{titlepage}

\baselineskip=20pt plus 2pt minus 2pt   

\newpage

          The subject of the present paper is the quantum
$R$-matrix
with the spectral parameter equal to zero (in  the  terms
 of  the
paper \cite{b1}). As is well known this $R$-matrix
satisfies the  Yang--
Baxter equation \cite{b2}
\begin{equation}
R_{12}R_{13}R_{23} = R_{23}R_{13}R_{12} \label{a1}
\end{equation}
Usually one considers $R$ as nondegenerate finite
dimensional matrix
acting on the tensor product
$V\otimes V$
of the finite dimensional vector space  $V$.
Then the notation
$R_{ij}$
in eq.(\ref{a1}) signifies the matrix on
$V \otimes V \otimes V$
acting as $R$ on the $i$-th and the $j$-th
components and as identity on
other component e.g.
$$ R_{12} = R \otimes 1\!\!\!\>{\rm I}$$

         The quantum $R$-matrix and the corresponding Yang--Baxter
equation  are  connected  with  the  various   problems   of   the
theoretical  and  mathematical  physics.  The   most   significant
achievement  is  the  observation  that  the  solutions   of   the
Yang-Baxter equation are related to the special algebras. These
are the deformed Lie algebras with comultiplication (Hopf
algebras) the so called "quantum groups" \cite{b3}. Other new
algebras introduced in \cite{b4} are based in the exponential
solutions of the Yang--Baxter equation. These  developments  have
led  to  the   construction   of   some generalized expressions
for $R$-matrix. However, it is  hardly  to
expect that  these  expressions  exhaust  all the solutions  of  the
Yang--Baxter  equation.  The   new   multiparametric   nonstandard
solution of (\ref{a1}) presented recently on \cite{b5} are one  such
example only.
         In the present  paper  we  would  like to reconsider  the
problem of the exact  solutions  of  Yang--Baxter  equation   from
slightly
different point of view. We are going to present a new  definition
for quantum $R$-matrix exploiting the idea  for  operator  acting
on  a functional space. For this aim let  us  consider  the  space
$M$  of functions of two arguments defined on the direct product
$G\times G$, where
$G$ is an arbitrary Lie group. The arbitrary function  from  $M$  have
the form
\begin{equation}
f\equiv f(h;g) \label{a2} \end{equation}
where the variables $h$ and $g$
run over all elements of group $G$  independently.  Now   we   can
give  the definition of the following (Right and Left) operators
\begin{eqnarray}
R^{R}f(h;g) & = & w^{R}(h;g)f(h;h^{-1}gh) \label{a3} \\
R^{L}f(h;g) & = & w^{L}(h;g)f(ghg^{-1};g) \label{a4}
\end{eqnarray}
Here the quantities $w^{R}(h;g)$
and $w^{L}(h;g)$
are so called "multiplicators" satisfying the equations
\begin{eqnarray}
w^{R}(h_{2};g)w^{R}(h_{1};h_{2}^{-1}gh_{2}) & = &
w^{R}(h_{2}h_{1};g)   \label{a5} \\
w^{L}(h;g_{2})w^{L}(g_{2}hg_{2}^{-1};g_{1}) & = &
w^{L}(h;g_{2}g_{1})   \label{a6}
\end{eqnarray}

      {\bf Theorem:} Let us denote with
$R^{A}_{ij} \;\;\;(A=R$ or $L$  and $i,j=1,2,3)$
the operators acting on $i$-th and $j$-th argument of the function
$f(g_{1};g_{2};g_{3})$
defined on $G\times G\times G $    as $R^{A}$
from definitions (\ref{a3}) or (\ref{a4}) respectively.
Then the operators
$R^{A}_{ij}$ for every fixed $A$
satisfy the Yang--Baxter equation (\ref{a1}).

         The proof of this  theorem is based on the properties
(\ref{a5})
and  (\ref{a6})
of the multiplicators. Although their explicit form is not  of
importance, we shall give a rather general
expression for these multiplicators. Let $T(g)$
is an arbitrary finite dimensional matrix  representation
of the group $G$. Then it is easy to verify that the quantity
\begin{equation}
w^{R}(h;g)=\left(\frac{{\rm Sp}KT(h^{-1}gh)}{{\rm Sp}KT(g)}\right)^d
 \label{a7}
\end{equation}
fulfil eq.(\ref{a5}), where with $K$
we have  denoted  an  arbitrary  but  fixed  matrix  of  the  same
dimension and $d$ is real constant.  We  can  write  down  similar
expression for the left multiplicator as well.

         Using the local coordinates in the Lie group G we can
give more effective form to our definitions,  which  is   fruitful
for the concrete applications. Here we shall choose the
coordinates  usually  called
"canonical  parameters"   \cite{b6}.   In   these   parameters   the
representations  of  the  local  Lie   group  $G$   have   strongly
exponential form. Let us suppose that $G$ has $n$  parameters  which
we denote with
$\alpha_{\mu},\beta_{\nu},\dots\;\;\;(\mu,\nu=1,2,\dots,n)$.
Then the function $f(h;g)$
from (\ref{a2})  can  be  written  in  the coordinate form as
functions depending on $2n$ variables $f(\alpha_{\mu};\beta_{\nu})$,
where $\alpha_{\mu} {\rm and} \beta_{\nu}$
are parameters of the group elements $h$ and $g$
respectively.

In what follows we will consider the right
$R$-matrix
from definition (~\ref{a3}) only. The corresponding
left matrices can be obtained analogously. If we denote with
$P^\nu_\mu(\alpha)$
the matrices of the adjoint representation of the group G, then the
parameters of element $h^{-1}gh$
have the form
\begin{equation}
\beta'_\mu = P_\mu^\nu(\alpha_\rho)\beta_\nu \label{a8}
\end{equation}
This expression follows directly from the definition of the adjoint
representation. Moreover, for the canonical parameters,
the following identity
\begin{equation}
P_\mu^\nu(\alpha_\rho)\alpha_\nu = \alpha_\mu  \label{a9}
\end{equation}
is fulfiled.
Then the definition (\ref{a3}) takes a new form
\begin{equation}
Rf(\alpha_\rho;\beta_\sigma) =
w(\alpha_\rho; \beta_\sigma)f(\alpha_\lambda;
P_\mu^\nu(\alpha_\tau)\beta_\nu)   \label{a10}
\end{equation}
Instead of the group elements in the arguments of the multiplicator  we
substitute their parameters. Then eq.(\ref{a5}) can be rewritten
in the form
\begin{equation}
w(\alpha_{2\rho}; \beta_\sigma)w(\alpha_{1\rho};
P_\mu^\nu(\alpha_{2\omega})\beta_\nu)=
w\left( m_\rho(\alpha_{2\omega};\alpha_{1\tau}); \beta_\sigma \right)
\label{a11}
\end{equation}
where with $m_\rho(\alpha_{2\omega};\alpha_{1\tau})$
we denote the parameters of the element $h_2h_1$.
It is clear that the functions
$m_\rho(\alpha_{2\omega};\alpha_{1\tau})$
express the group's multiplication law. In terms of the
group  parameters we have chosen,
these   functions   satisfy the following simple chain of
identities.
\begin{equation}
m_\rho(\beta_\mu;\alpha_\nu)=
m_\rho(\alpha_\nu; P_\mu^\sigma(\alpha)\beta_\sigma)=
m_\rho\left(\left(P^{-1}\right)^\nu_\mu(\beta)\alpha_\nu;
\beta_\sigma\right)
\label{a12}
\end{equation}

             In general the above defined here $R$-matrix is
 an infinite-dimensional . As is well known for the  Lie
groups,  the functions
$P^\nu_\mu(\alpha)\beta_\nu, w(\alpha_\rho;\beta_\sigma)$
and $m_\rho(\alpha_\mu;\beta_\nu)$
are smooth. According to the definition (\ref{a10}) of the  operator $R$
the subspace $S$ of $M$ consisting  of  smooth  functions  is
invariant under the action of this  operator.  Using the basis  of
orthonormal functions in the subspace $S$ we can obtain  the  matrix
form of the operator $R$.

Another invariant subspace one can obtain when $G$  is  an  simple
compact group with structure constants ${C^{\mu\nu}}_\rho$. It  is
well known that in this case
there exists a positive defined scalar  product  (Killing  metric)
invariant with respect to the
adjoint representation of this group. In
particular if with $\eta^{\mu\nu}$
is the Killing metric tensor, then the scalar square
\begin{equation}
\beta^2 = \eta^{\mu\nu}\beta_\mu\beta_\nu  \label{a13}
\end{equation}
is invariant under the transformation (\ref{a8})

{\bf Remark:} The Killing metric tensor is defined as follows:
\begin{equation}
\eta^{\mu\nu} = \frac{-1}{n-1}
{C^{\mu\rho}}_\sigma {C^{\nu\sigma}}_\rho    \label{a17}
\end{equation}

         Let us consider now the  special  subspace $D$ of $M$
constructed from the functions having the following form:
\begin{equation}
l(\alpha_\mu;\beta_\nu)\delta(\alpha^2-p^2)
\delta(\beta^2-p^2)\delta\left((\alpha\beta)-q\right)
\label{a14}
\end{equation}
where $p$
and   $q$
are arbitrary real constants and the functions $l(\alpha_\mu;\beta_\nu)$
are smooth. Here we have used the following notation:
$$ (\alpha\beta) = \eta^{\mu\nu}\alpha_\mu\beta_\nu.$$

         It is easy to verify that the  subspace $D$ of  functions
(\ref{a14}) is invariant under the  action  of  the  operator $R$ from
(\ref{a10}). In this check we have to use  the  property
(\ref{a9}) in order to convince ourselves
that the quantity $(\alpha\beta)$
in  the  delta-function  argument  is  not   changed   too   (some
definitions used here are given in the Appendix).

         In our opinion the  main  question  now  is  whether  a
finite dimensional $R$-matrix can be obtained from our  definitions.
To answer it we have to investigate the  structure  of  the
space $M$ from the point of view of the operator  action (\ref{a10}).
It is easy to understand that the finite dimensional $R$-matrix will be
defined on a finite dimensional subspace of $M$ invariant under  the
action of the operator $R$. In the general case
the problem to find such subspaces is difficult enough.
However, when $G$ is a simple compact group this  problem  can  be
solved in any particular case using the group invariants. In  what
follows we are going to demonstrate how  to  separate  the  finite
dimensional subspace from $D$ in the case when $G$  coincide  with
$SU(2)$. Then the structure constants have the form:
$${C^{\mu\nu}}_\rho = \epsilon_{\mu\nu\rho},$$
where $\epsilon_{\mu\nu\rho}$ are the components of the completely
antisymmetric tensor for $n=3$. The Killing metric tensor coincide
with unit tensor
$$\eta_{\mu\nu} = \delta_{\mu\nu}$$
Using the definition of the adjoint representation one can derive
the following identity
\begin{equation}
P_\mu^\rho(\beta)\delta_{\rho\nu} =
P_\mu^{\mu'}(\alpha) P_\nu^{\nu'}(\alpha) \delta_{\nu'\nu''}
P_{\mu'}^{\nu''}\left(P_\rho^\sigma(\alpha)\beta_\sigma\right)
\label{a18}
\end{equation}
which means that the matrix $P_\mu^\nu(\beta)$
is an invariant tensor field of second rank. Then the
general expression of $P_\mu^\nu(\alpha)$
has the form:
\begin{equation}
P_\mu^\nu(\alpha) = A(\alpha^2)\delta_{\mu\nu} +
B(\alpha^2)\epsilon_{\nu\rho\mu}\alpha_\rho +
E(\alpha^2)\delta_{\nu\nu'}\alpha_\mu\alpha_{\nu'}
\label{a19}
\end{equation}
All the three functions of the invariant  $\alpha^2$
can be found using the condition of the invariance of the  Killing
metric tensor:
\begin{equation}
\delta_{\mu\nu}P_\mu^\rho(\alpha)P_\nu^\sigma(\alpha) =
\delta_{\rho\sigma}
\label{a20}
\end{equation}
and the eq.(\ref{a9}). The result is
\begin{equation}
P_\mu^\nu(\alpha) = \cos{\sqrt{\alpha^2}}\delta_{\mu\nu} +
\frac{\sin{\sqrt{\alpha^2}}}{\sqrt{\alpha^2}}
\epsilon_{\mu\nu\rho}\alpha_\rho +
\frac{2\sin{\frac{\sqrt{\alpha^2}}{2}}}{\alpha^2}
\delta_{\nu\nu'}\alpha_\mu\alpha_{\nu'}
\label{a21}
\end{equation}
The same we can obtain calculating the matrix
\begin{equation}
P_\mu^\nu(\alpha) = \exp{i\{ \epsilon_{\nu\rho\mu}\alpha_\rho\} }
\label{a22}
\end{equation}
as well known power series.

         Now we are  going  to  consider  the definition (\ref{a10})
without the multiplicator (i.e. we choose $d=0$ from eq.(\ref{a7})) and
with  the  functions taken from  the  subspace  $D$.  Then   instead
eq.(\ref{a10}) we have
\begin{equation}
R l(\alpha_\mu;\beta_\nu)
\delta(\alpha^2-p^2) \delta(\beta^2-p^2)
\delta\left( (\alpha\beta)-q\right) =
l(\alpha_\mu; P_\rho^\sigma(\alpha)\beta_\sigma)
\delta(\alpha^2-p^2) \delta(\beta^2-p^2)
\delta\left( (\alpha\beta)-q\right)
\label{a23}
\end{equation}
Taking into account the explicit form (\ref{a19}) or (\ref{a21}) of
$P_\mu^\nu(\alpha)$ we can see that the quantity
$P_\mu^\nu(\alpha)\beta_\nu$
entering in eq.(\ref{a23}) takes the following simple form:
\begin{equation}
P_\mu^\nu(\alpha)\beta_\nu =
a_1\alpha_\mu + a_2\beta_\mu +
a_3 {C^{\rho\sigma}}_\mu\alpha_\rho\beta_\sigma
\label{a24}
\end{equation}
Because of the delta-functions in eq.(\ref{a23}) the quantities
$a_1, a_2$ and $a_3$
are constants expressed quite definitely by  $p$ and $q$.

         Now we can see that
all the functions $l(\alpha_\mu;\beta_\nu)$
which coincide with the polynomials of the following type
\begin{equation}
l(\alpha_\mu;\beta_\nu) = b_0 +
b_1^\mu\alpha_\mu + b_2^\mu\beta_\mu +
b_3^\mu {C^{\rho\sigma}}_\mu\alpha_\rho\beta_\sigma
\label{a25}
\end{equation}
form a finite dimensional subspace of $D$.
The action of the operator
(\ref{a23}) on these functions leads to a linear  changing  of  the
polynomial coefficients only. The above
mentioned finite dimensional
$R$-matrix  can  be  extracted  as  the corresponding constant matrix
of this linear transformation of the polynomial coefficients. This
procedure coincide with the standard one, used for separation  the
finite dimensional representation from the  general  functional
definition of the infinite dimensional operator representation  of
a Lie group. That is why it is not necessary to  consider  this
procedure in more detail.

\vskip 20pt
  {\bf Acknowledgment:} The author is grateful to Drs.  S.Pacheva
and E.Nissimov for useful discussions and  to  Dr.  M.Stoilov  for
help at the preparation of this article.

\vskip 20pt
{\bf Appendix}

         Let $G$ be a $n$-parametric Lie group and
$\alpha_\mu, \;\;\;(\mu=1,2,\dots,n)$
be the canonical parameters of element $g\in G$.
According to \cite{b6} the canonical parameters  are  defined
with the help of the group structure functions
$S_\mu^\nu(\alpha_\rho)$
in the following way
\begin{equation}  \label{A1}
S_\mu^\nu(\alpha_\rho)\alpha_\nu = \alpha_\mu
\end{equation}

         Let $T(g)$
be an arbitrary representation of the group $G$ with generators
$I^\mu$.
Then the adjoint representation matrices are defined as follows
\begin{equation}  \label{A2}
T^{-1}(g)I^\mu T(g) = I^\nu P_\nu^\mu(\alpha)
\end{equation}

         The main consequence from eq.(\ref{A1}) is that operator
$T(g)$ can be written in the form:
\begin{equation}  \label{A3}
T(g) =\exp{\{i I^\mu\alpha_\mu\} }
\end{equation}
Combining the formulas (\ref{A2}) and (\ref{A3}) we obtain that
\begin{equation}  \label{A4}
\exp{\{i I^\mu\beta_\mu\} }
\exp{\{i I^\nu\alpha_\nu\} } =
\exp{\{i I^\nu\alpha_\nu\} }
\exp{\{i I^\rho P_\rho^\sigma(\alpha)\beta_\sigma\} }
\end{equation}
from which the eq.(\ref{a12}) follows  immediately. The
case when $\alpha_\mu \equiv \beta_\mu$
the eq.(\ref{A4}) leads to the identity (\ref{a9}).

 Finally the identity (\ref{a18}) also follows from eq.(\ref{A4})
when the later is written for the adjoint representation i.e. when
the matrices of the generators $I^\mu$
are $(I^\mu)_\rho^\sigma = {C^{\mu\sigma}}_\rho \;\;
({C^{\mu\nu}}_\rho$
are the structure constant of group $G$ as above).

\end{document}